\RequirePackage[displaymath]{lineno} 
\documentclass[aps,twocolumn,showpacs,byrevtex,prl,reprint]{revtex4-1}
\usepackage{epsfig}
\usepackage{graphicx}
\usepackage{dcolumn}
\usepackage{bm}
\usepackage{overpic}
\usepackage{subfigure}
\usepackage{float}
\usepackage{color}
\usepackage{amsmath}
\usepackage{mathcomp}
\usepackage{mathrsfs}
\usepackage{multirow}
\usepackage{rotating}
\usepackage{hyperref}
\usepackage{threeparttable}
\usepackage[outdir=./]{epstopdf}

\begin{document}
\normalsize
\parskip=5pt plus 1pt minus 1pt

\title{ \boldmath Observation of $D \to a_{0}(980)\pi$ in the decays $D^{0} \rightarrow \pi^{+}\pi^{-}\eta$ and $D^{+} \rightarrow \pi^{+}\pi^{0}\eta$  }
\vspace{-1cm}

\author{
   \begin{small}
    \begin{center}
    M.~Ablikim$^{1}$, M.~N.~Achasov$^{4,c}$, P.~Adlarson$^{75}$, O.~Afedulidis$^{3}$, X.~C.~Ai$^{80}$, R.~Aliberti$^{35}$, A.~Amoroso$^{74A,74C}$, Q.~An$^{71,58,a}$, Y.~Bai$^{57}$, O.~Bakina$^{36}$
% ...
         \\
         \vspace{0.2cm}
   (BESIII Collaboration)\\
\vspace{0.2cm} {\it 
$^{1}$ Institute of High Energy Physics, Beijing 100049, People's Republic of China\\
% ...
$^{3}$ Bochum Ruhr-University, D-44780 Bochum, Germany\\
$^{4}$ Budker Institute of Nuclear Physics SB RAS (BINP), Novosibirsk 630090, Russia\\
% ...
$^{35}$ Johannes Gutenberg University of Mainz, Johann-Joachim-Becher-Weg 45, D-55099 Mainz, Germany\\
$^{36}$ Joint Institute for Nuclear Research, 141980 Dubna, Moscow region, Russia\\
% ...
$^{57}$ Southeast University, Nanjing 211100, People's Republic of China\\
$^{58}$ State Key Laboratory of Particle Detection and Electronics, Beijing 100049, Hefei 230026, People's Republic of China\\
% ...
$^{71}$ University of Science and Technology of China, Hefei 230026, People's Republic of China\\
% ...
$^{74}$ University of Turin and INFN, (A)University of Turin, I-10125, Turin, Italy; (B)University of Eastern Piedmont, I-15121, Alessandria, Italy; (C)INFN, I-10125, Turin, Italy\\
$^{75}$ Uppsala University, Box 516, SE-75120 Uppsala, Sweden\\
% ...
$^{80}$ Zhengzhou University, Zhengzhou 450001, People's Republic of China\\
\vspace{0.2cm}
$^{a}$ Deceased\\
% ...
$^{c}$ Also at the Novosibirsk State University, Novosibirsk, 630090, Russia\\
% ...
}\end{center}
\vspace{0.4cm}
\end{small}
}

\affiliation{}
\vspace{-4cm}

\begin{abstract}
We report the first amplitude analysis of the decays $D^{0} \to \pi^{+} \pi^{-} \eta$ and $D^{+} \rightarrow \pi^{+}\pi^{0}\eta$ using a data sample taken with the BESIII detector at the center-of-mass energy of 3.773~GeV, corresponding to an integrated luminosity of 7.9 ${\rm fb}^{-1}$.  
The contribution from the process $D^{0(+)} \to a_{0}(980)^{+} \pi^{-(0)}$ is significantly larger than the $D^{0(+)} \to a_{0}(980)^{-(0)} \pi^{+}$ contribution.
The ratios  $\mathcal{B}(D^{0} \rightarrow a_{0}(980)^{+}\pi^{-})/\mathcal{B}(D^{0} \rightarrow a_{0}(980)^{-}\pi^{+})$ and $\mathcal{B}(D^{+} \rightarrow a_{0}(980)^{+}\pi^{0})/\mathcal{B}(D^{+} \rightarrow a_{0}(980)^{0}\pi^{+})$ are measured to be $7.5^{+2.5}_{-0.8\,\mathrm{stat.}}\pm1.7_{\mathrm{syst.}}$ and $2.6\pm0.6_{\mathrm{stat.}}\pm0.3_{\mathrm{syst.}}$, respectively. 
The measured $D^{0}$ ratio disagrees with the theoretical predictions by orders of magnitudes, thus implying a substantial contribution from final-state interactions.

\end{abstract}
\maketitle
Theoretical predictions of the strong interaction in the charm sector are challenging, since quantum chromodynamics calculations involve non-perturbative contributions.
The W-annihilation (WA) and W-exchange (WE) processes, which are strictly suppressed in $B$-meson decays, 
can occur in $D$ decays as a result of final-state interactions (FSI).  
They are expected to be dominated by non-perturbative effects, which leads to significant uncertainties when making theoretical predictions,  
since these effects  depend strongly on the cutoff values and the unknown phases between different processes~\cite{Cheng:2010ry,Qin:2013tje}.
Therefore, the study of decays with a significant contribution from WA or WE processes represents a promising method for investigating the dynamics of charm decays.

The BESIII Collaboration has observed the pure WA decays $D_{s}^{+} \to a_{0}(980)^{+(0)} \pi^{0(+)}$~\cite{BESIII:2019jjr},
which indicates a sizeable contribution from FSI in the WA processes for the $D \to SP$ sector (where $S$ and $P$ denote scalar and pseudo-scalar particles, respectively).
Theorists have explained the observed large WA amplitude, as well as the  amplitude symmetry $A(D_{s}^{+} \to a_0(980)^{+}\pi^{0}) = -A(D_{s}^{+} \to a_0(980)^{0}\pi^{+})$,
taking into account the contribution from the $D_{s}^{+} \to \rho^{+}\eta$ and $D_{s}^{+} \to \bar{K}^{*}(892)^{0}K^{+}(K^{*}(892)^{+}\bar{K}^{0})$ decays, since they exhibit large
branching fractions (BF) and involve the WA amplitude process at the quark level~\cite{Hsiao:2019ait,Ling:2021qzl}. 
This behavior supports the interpretations of $a_{0}(980)$ as a tetraquark or a molecular state~\cite{Ling:2021qzl}. 
More recently, further measurements involving $D_{s}^{+}$ decays have been performed, in particular those that have led to 
the observation of the $a_{0}(1817)^{+(0)}$ resonance~\cite{BESIII:2021anf,BESIII:2022npc}, which is expected as an excited state of the $a_{0}(980)^{+(0)}$~\cite{Guo:2022xqu}, support
the interpretation of  these two resonances as $K^{(*)}\bar{K}^{(*)}$ molecules ~\cite{Wang:2021jub,Oset:2023hyt}.  

In $D^{0}$ decays,  the relative ratio 
$r_{+/-} = \mathcal{B}(D^{0} \to a_{0}(980)^{+} \pi^{-})/ \mathcal{B}(D^{0} \to a_{0}(980)^{-} \pi^{+})$ is expected to be less than 0.05~\cite{Cheng:2022vbw}, when ignoring the WE process.
Until now, attempts to measure this ratio have not been conclusive; both the CLEO and LHCb Collaborations have studied the $a_{0}(980)^{\pm} \pi^{\mp}$ contributions to the  decays $D^{0} \to K_{S}^{0}K^{\pm}\pi^{\mp}$~\cite{CLEO:2012obf,LHCb:2015lnk}, but with large uncertainties;
the Belle Collaboration has studied  $D^{0} \to \pi^{+}\pi^{-}\eta$ decays~\cite{Belle:2021dfa}, and has only observed the $ a_{0}(980)$ peak in the $M(\pi^{+}\eta)$ projection.

In analogy to what is observed in $D_{s}^{+} \to a_{0}(980)^{+(0)} \pi^{0(+)}$ decays~\cite{BESIII:2019jjr}, 
sizeable contributions from FSI are expected to enhance the WA process in the corresponding $D^{+}$ decays.
However, in this case the symmetry observed in $D_{s}^{+}$ decays is expected to be violated, since further short-distance contributions must be considered, 
which can be expressed as a color-allowed external $W$-emission tree (T) diagram and a color-suppressed external $W$-emission tree diagram in the topological diagram approach~\cite{Cheng:2022vbw}. The measurement of the BFs for $D^{0}$ ($D^{+}$) decays to $a_{0}(980)\pi$ and of the corresponding relative ratios $r_{+/-}$ ( 
$r_{+/0} = \mathcal{B}(D^{+} \to a_{0}(980)^{+} \pi^{0})/ \mathcal{B}(D^{+} \to a_{0}(980)^{0} \pi^{+})$) 
can constrain the size and phase of the amplitude of the WE (WA) process and improve the knowledge about the role the $a_{0}(980)$ plays in charm decays~\cite{Ikeno:2021kzf}.  

In this Letter, we perform amplitude analyses of  $D^{0(+)} \to \pi^{+}\pi^{-(0)} \eta$ decays, to study the contributions from the intermediate processes $D^{0(+)} \to a_{0}(980)^{+} \pi^{-(0)}$, $D^{0(+)} \to a_{0}(980)^{-(0)} \pi^{+}$ and $D^{0(+)} \to \rho^{0(+)} \eta$. The analyses are based on $e^+e^-$ collision data recorded with the BESIII detector at the center-of-mass energy of 3.773 GeV, corresponding to an integrated luminosity of 7.9~${\rm fb}^{-1}$. Charge conjugation is implied throughout this Letter, as well as the equivalences $a_{0}(980)^{\pm(0)} \to \pi^{\pm(0)} \eta$ and $\mathcal{B}(a_{0}(980)^{+} \to \pi^{+}\eta) = \mathcal{B}(a_{0}(980)^{0} \to \pi^{0}\eta)$.

A detailed description of the BESIII detector design and performance can be found in Ref.~\cite{detector}. About 63$\%$  of the data analyzed in this Letter profits from an upgrade of the end-cap time-of-flight system with multi-gap resistive plate chambers with a time resolution of 60~ps~\cite{MRPC}. Simulated data samples produced with a {\sc geant4}-based~\cite{sim} Monte Carlo (MC) package, which
includes the geometric description of the BESIII detector and the detector response, are used to determine detection efficiencies
and to estimate backgrounds. The simulation models the beam-energy spread and initial-state radiation (ISR) in $e^+e^-$ annihilations with the generator {\sc kkmc}~\cite{ref:kkmc}. 
The inclusive MC sample includes the production of $D\bar{D}$ pairs (including quantum coherence for the neutral $D$ channels),
the non-$D\bar{D}$ decays of the $\psi(3770)$, the ISR production of the $J/\psi$ and $\psi(3686)$ states, and the
continuum processes incorporated in {\sc kkmc}~\cite{ref:kkmc}.

The charged-track selection, particle identification (PID), $K_{S}^{0}$, $\pi^{0}$ and $\eta$ reconstruction use the same criteria described in Ref.~\cite{BESIII:2019jjr},
except for the invariant-mass window around the $\eta$, which is set to $0.45<M(\gamma\gamma)_{\eta}<0.55$~GeV$/c^{2}$.
The $D$ mesons are identified using the beam-constrained mass $M_{\mathrm{BC}} = \sqrt{E_{\mathrm{beam}}^{2} - |\Vec{P}_{D}|^{2}}$ 
and the deviation of the reconstructed energy from the expected energy $\Delta E = E_{D} - E_{\mathrm{beam}}$, where $(E_{D},\Vec{P}_{D})$ is the four-momentum of the $D$ meson, and $E_{\mathrm{beam}}$ is the beam energy. 
A double-tag (DT) technique~\cite{MARK-III:1985hbd} is employed to suppress the background. 
On both the tag and the signal side, any candidate with $M_{\mathrm{BC}}<1.83$~GeV$/c^{2}$ or $|\Delta E|>0.1$~GeV is rejected; 
if multiple combinations survive in an event, the one with the $M_{\mathrm{BC}}$ closest to the known $D$ meson mass from Particle Data Group (PDG)~\cite{Workman:2022ynf} is retained. 

For the $D^{0}$ channel, four tag modes ($\bar{D}^{0} \to K^{+} \pi^{-}$, $\bar{D}^{0} \to K^{+} \pi^{-} \pi^{0} (\pi^{0})$
and $\bar{D}^{0} \to K^{+} \pi^{-} \pi^{-} \pi^{+}$) are used, while for the $D^{+}$ channel we use six tag modes ($D^{-} \to K^{+}\pi^{-}\pi^{-} (\pi^{0})$, $D^{-} \to K^{0}_{S}\pi^{-} (\pi^{0})$, $D^{-} \to K^{0}_{S}\pi^{-}\pi^{-}\pi^{+}$ and $D^{-} \to K^{+}K^{-}\pi^{-}$).
Signal MC samples with $\psi(3770) \to D\bar{D}$, $\bar{D} \to tag~modes$ and $D \to signal~modes$ are produced, in which the signal decays 
$D^{0(+)} \to \pi^{+} \pi^{-(0)} \eta$ are generated with the amplitude models that result from the studies presented in this Letter. 
For the tag channels, the $M_{\mathrm{BC}}$ signal windows are set to be $\pm 6$~MeV$/c^{2}$ around the known $D$ mass~\cite{Workman:2022ynf}, 
while the $\Delta E$ signal windows are set to be 3.5 times the $\Delta E$ resolution around the fitted peak. 

On the signal side, we select  $D^{0}$ ($D^{+}$) candidates with $M_{\mathrm{BC}}$ within $[1.858,~1.874]$~GeV$/c^{2}$ ($[1.860,~1.880]$~GeV$/c^{2}$).
Furthermore, for the $D^{0}$ channel, the requirement $|M(\pi^{+}\pi^{-}) - m(K_{S}^{0})|>0.03$~GeV$/c^{2}$ is imposed to remove the peaking background from the $D^{0} \to K_{S}^{0}\eta$ decays, where $m(K_{S}^{0})$ is the known $K_{S}^{0}$ mass~\cite{Workman:2022ynf}.
Since the dominant background originates from wrong $\eta \to \gamma \gamma$ candidates, 
a multivariate analysis (MVA)~\cite{Hocker:2007ht} is performed to select events for use in the amplitude analysis. 
This MVA involves the development of a Gradient Boosted Decision Tree (BDTG) classifier based on the inclusive MC sample, which operates on three discriminating variables: the $\gamma\gamma$ invariant mass $M(\gamma\gamma)_{\eta}$, the goodness of the kinematic fit constraining the $\gamma\gamma$ invariant mass to the known $\eta$ mass $\chi^{2}(\eta)$, and the helicity angle of the higher energy photon 
from the $\eta$ decay. A requirement on the BDTG output is imposed,  
which retains 83\% (77\%) of the signal and rejects 78\% (84\%) of the background for the $D^{0}$ ($D^{+}$) channel according to studies performed with MC simulation.
Additionally, the selection 
$|\Delta E|<0.045$~GeV ($|\Delta E|<0.040$~GeV) for the $D^{0}$ ($D^{+}$) channel is applied.
The final sample contains 1678 (1226) $D^{0} (D^{+}) \to \pi^{+} \pi^{-} (\pi^{0}) \eta$ candidates with a purity of $(74.1\pm1.2)\%$ ($(65.7\pm1.7)\%$).

The amplitude analysis is performed on the accepted candidate events with an unbinned maximum-likelihood fit. The logarithm of the likelihood is constructed as 
\begin{eqnarray}
\begin{aligned}
\ln L = \ln(f_{s}\Tilde{S}(p) + (1-f_{s})\Tilde{B}(p)),
\end{aligned}
\end{eqnarray}
where $f_{s}$ is the signal purity, $p$ is the four-momenta of final particles, $\Tilde{S}(p)$ and $\Tilde{B}(p)$ are the signal and background probability density function (PDF), expressed as 
\begin{eqnarray}
\begin{aligned}
\Tilde{S}(p)& = \frac{\epsilon(p)|\mathcal{M}(p)|^{2}R_{3}(p)}{\int{\epsilon(p)|\mathcal{M}(p)|^{2}R_{3}(p)\mathrm{d}p}},\\
\Tilde{B}(p)& = \frac{\epsilon(p)B_{\epsilon}(p)R_{3}(p)}{\int{\epsilon(p)B_{\epsilon}(p)R_{3}(p)\mathrm{d}p}}.	
\end{aligned}
\end{eqnarray}
Here, $R_{3}(p)$ is the three-body phase-space factor, $\epsilon(p)$ is the efficiency function, $\mathcal{M}(p)$ is the signal amplitude, $B(p)$ is the background shape, and  $B_{\epsilon}(p) = B(p)/\epsilon(p)$ is the efficiency-corrected background  shape.
For the $D^{0}$ channel,  the $B(p)$ term is extracted from the $\Delta E$ sideband region ($0.05<|\Delta E|<0.10$~GeV);  for the $D^{+}$ channel, since the resolution is much wider than for the $D^{0}$, the inclusive MC sample is used.
The total signal amplitude $\mathcal{M}(p)= \sum_{\alpha}{c_{\alpha}e^{i\phi_{\alpha}}A_{\alpha}}$ is modeled as the coherent sum of the amplitudes of all the intermediate processes, where $c_{\alpha}$ and $\phi_{\alpha}$ are the magnitude and phase of the $\alpha^{\mathrm{th}}$ amplitude, respectively, and the $\alpha^{\mathrm{th}}$ amplitude is given by $A_{\alpha} = P_{\alpha}S_{\alpha}F^{r}_{\alpha}F^{D}_{\alpha}$. Here, $P_{\alpha}$ is the propagator, generally following the relativistic Breit-Wigner formula except for the $\rho$ and $a_{0}(980)$ states. For the $\rho^{+}$ we use the GS formula~\cite{GS}, and for the $\rho^{0}$ description $\rho-\omega$ mixing is additionally considered~\cite{BESIII:2019ymv}. When modeling the $a_{0}(980)$ the two channel-coupled Flatt\'e formula $P_{a_{0}(980)} = 1/[(m_{0}^{2} - s_{a}) - i(g_{\eta\pi}^{2}\rho_{\eta\pi} + g_{K\bar{K}}^{2}\rho_{K\bar{K}})]$ is used, with $g_{\pi\eta}$ ($g_{K\bar{K}}$) and $\rho_{\pi\eta}$ ($\rho_{K\bar{K}}$) representing the coupling constant from Ref.~\cite{BESIII:2016tqo} and the phase-space factor $q/\sqrt{s_{a}}$, respectively, where $s_{a}$ is the invariant-mass squared of the $a_{0}(980)$ candidate and $q$ is the total momentum of the daughter particles in the $a_{0}(980)$ rest frame. For $\pi^{+}\pi^{-}$ $S$-wave scattering, the K-matrix formalism~\cite{Anisovich:2002ij} is used, with parameters taken from Ref.~\cite{LHCb:2015klp}.
The $S_{\alpha}$ term is the spin factor and is constructed with the covariant-tensor formalism~\cite{Zou:2002ar}.
The $F^{r}_{\alpha}$ ($F^{D}_{\alpha}$) term is the barrier factor for the intermediate state (the $D$ meson)~\cite{BESIII:2022npc}.
The relative contribution of the $\alpha^{\mathrm{th}}$ amplitude to the $\beta^{\mathrm{th}}$ amplitude is quantified by the ratio $r_{\alpha/\beta} = \int{|c_{\alpha}A_{\alpha}(p)|^{2}R_{3}(p)\mathrm{d}p}/\int{|c_{\beta}A_{\beta}(p)|^{2}R_{3}(p)\mathrm{d}p}$. When substituting $c_{\beta}A_{\beta}$ with $\mathcal{M}(p)$, the ratio becomes the fit fraction of the $\alpha^{\mathrm{th}}$ amplitude ($\mathrm{FF}_{\alpha}$).

\begin{figure}[htbp]
\begin{center}
\begin{minipage}[b]{0.23\textwidth}
\epsfig{width=0.98\textwidth,clip=true,file=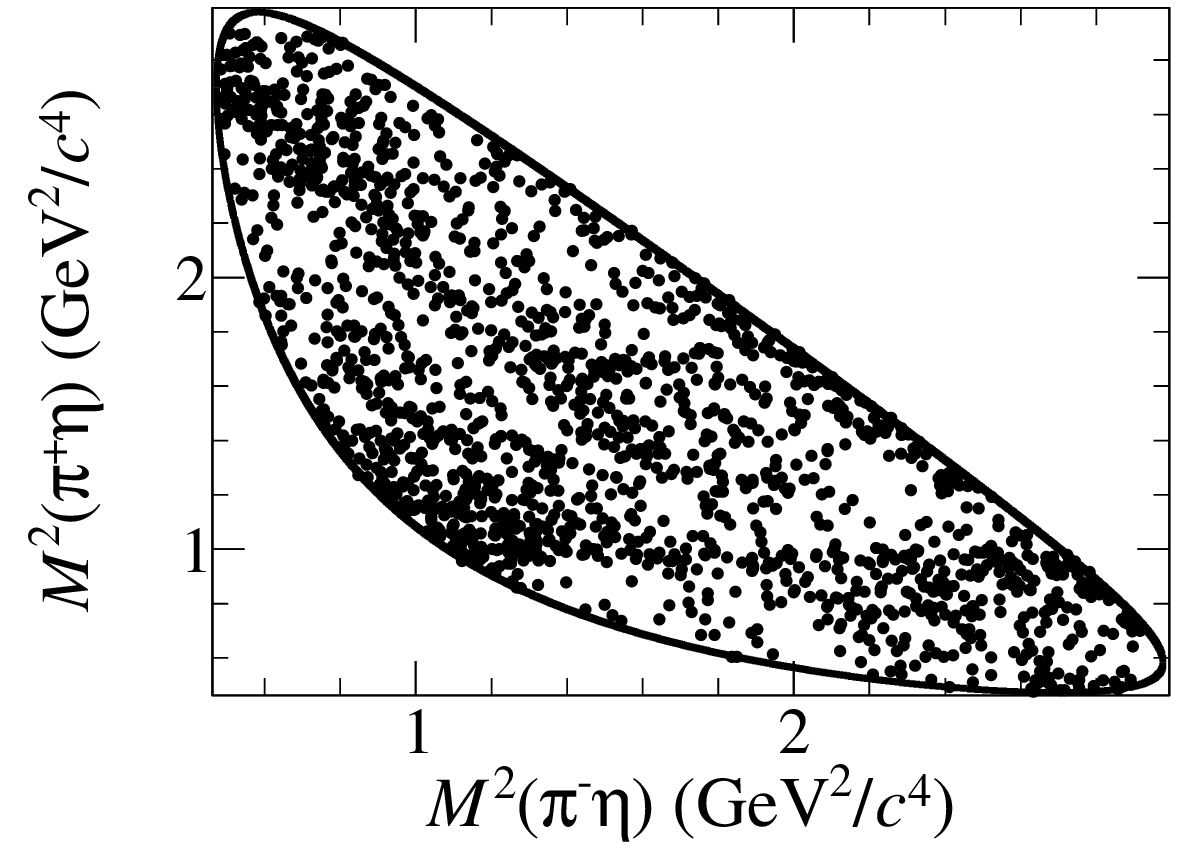}
\put(-25,65){(a)}
\end{minipage}
\begin{minipage}[b]{0.23\textwidth}
\epsfig{width=0.98\textwidth,clip=true,file=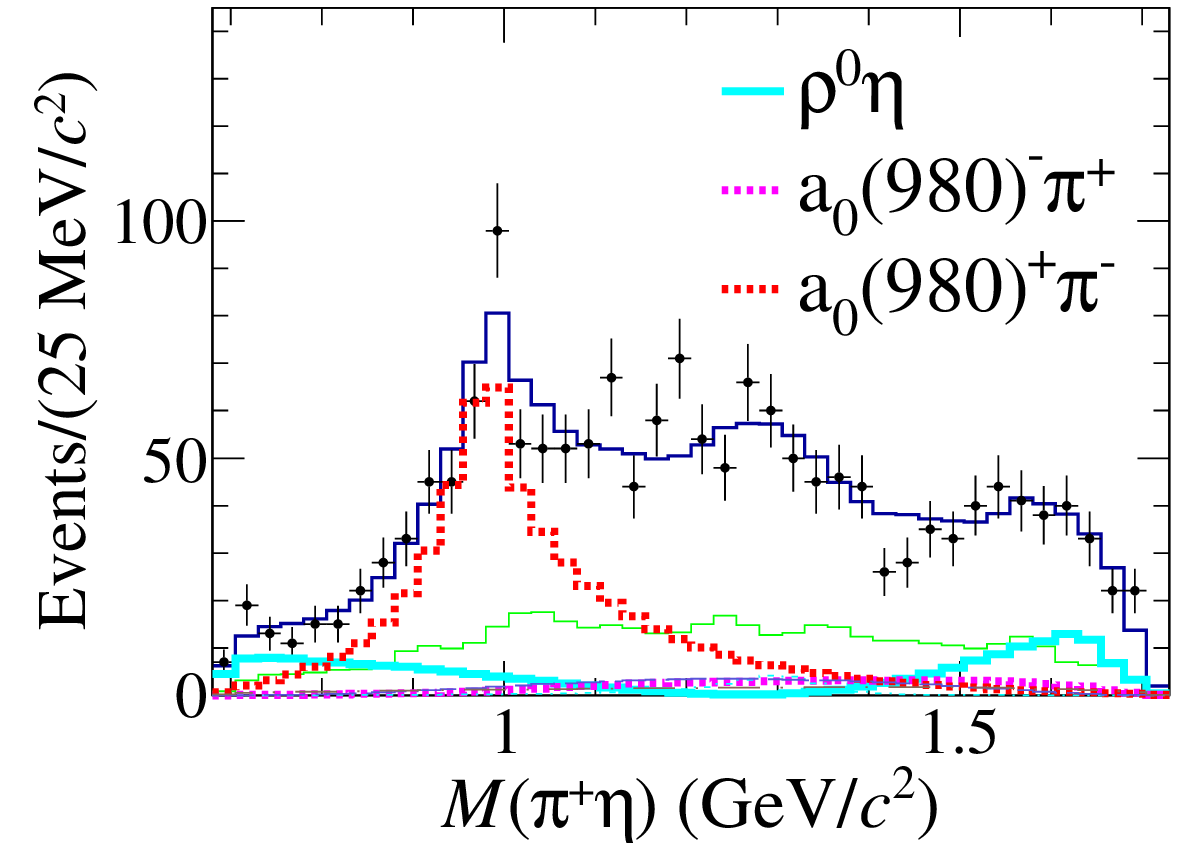}
\put(-85,65){(b)}
\end{minipage}
\begin{minipage}[b]{0.23\textwidth}
\epsfig{width=0.98\textwidth,clip=true,file=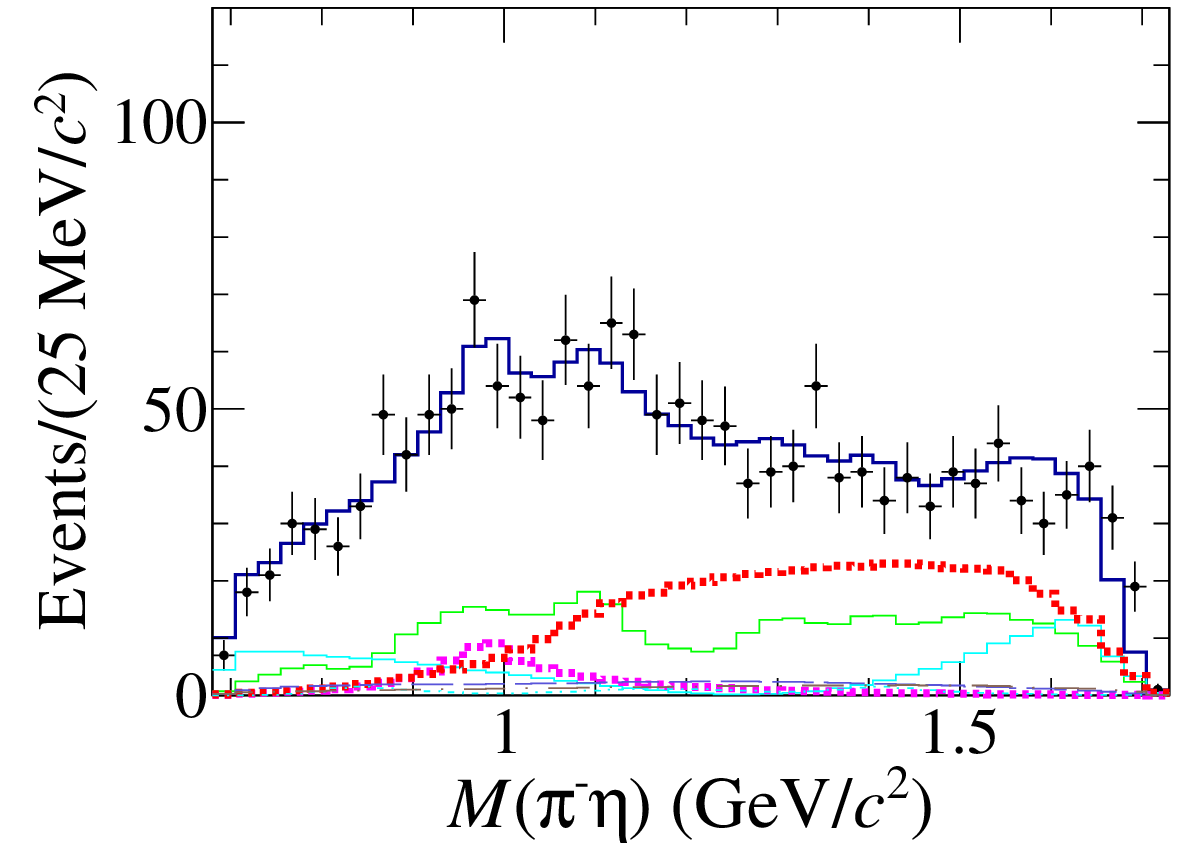}
\put(-85,65){(c)}
\end{minipage}
\begin{minipage}[b]{0.23\textwidth}
\epsfig{width=0.98\textwidth,clip=true,file=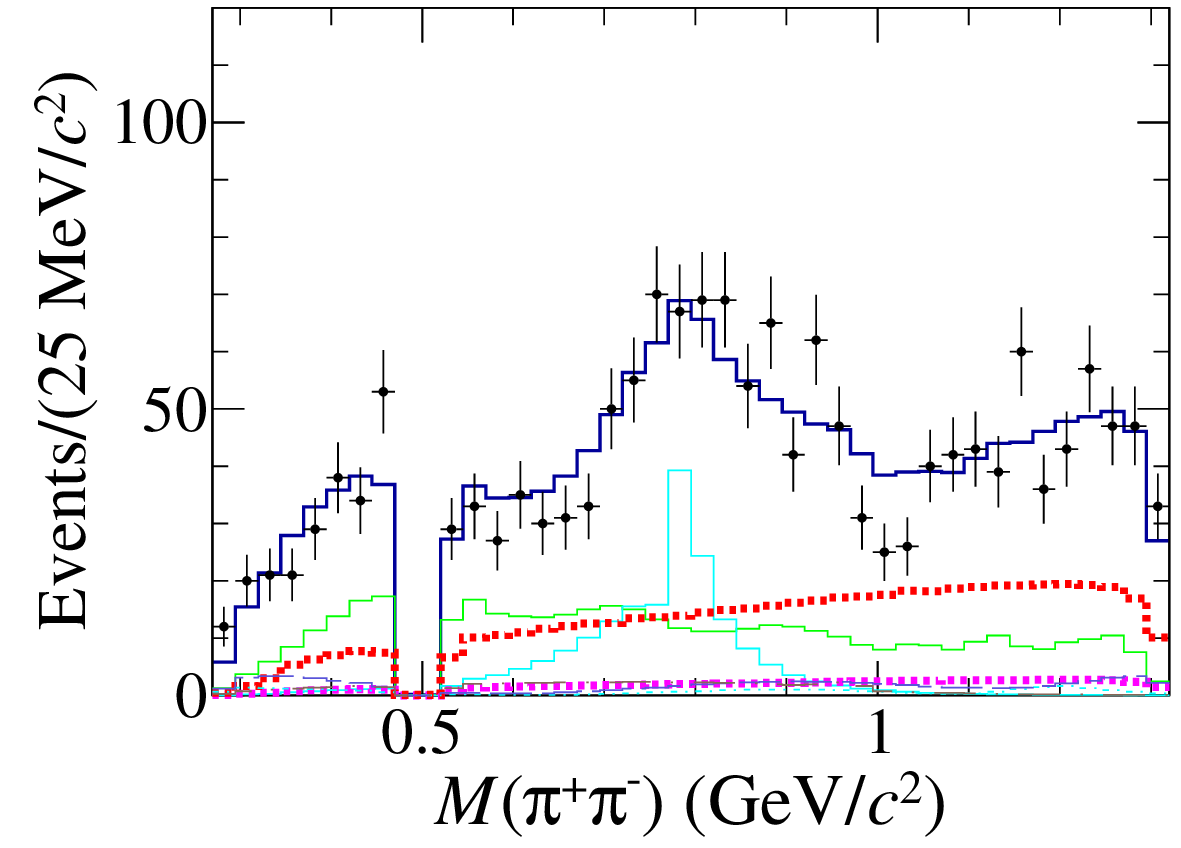}
\put(-85,65){(d)}
\end{minipage}
\begin{minipage}[b]{0.23\textwidth}
\epsfig{width=0.98\textwidth,clip=true,file=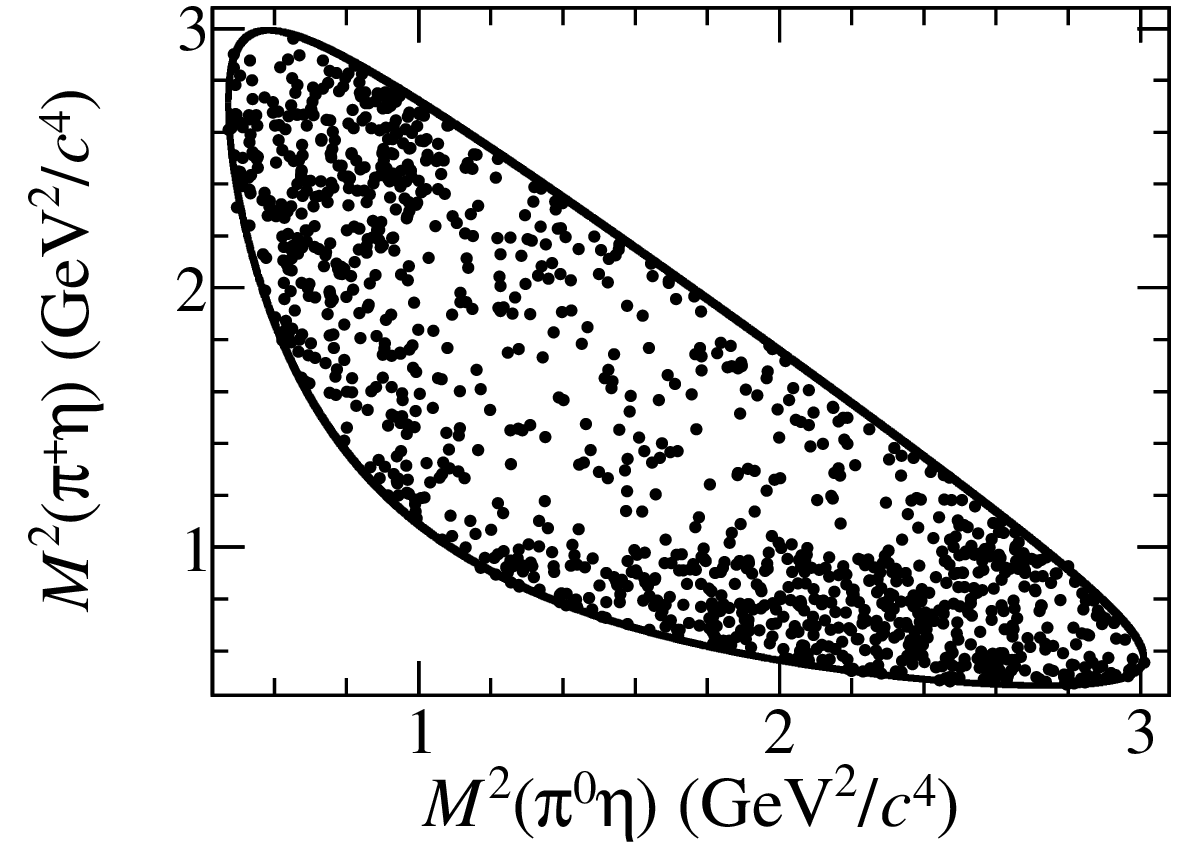}
\put(-25,65){(e)}
\end{minipage}
\begin{minipage}[b]{0.23\textwidth}
\epsfig{width=0.98\textwidth,clip=true,file=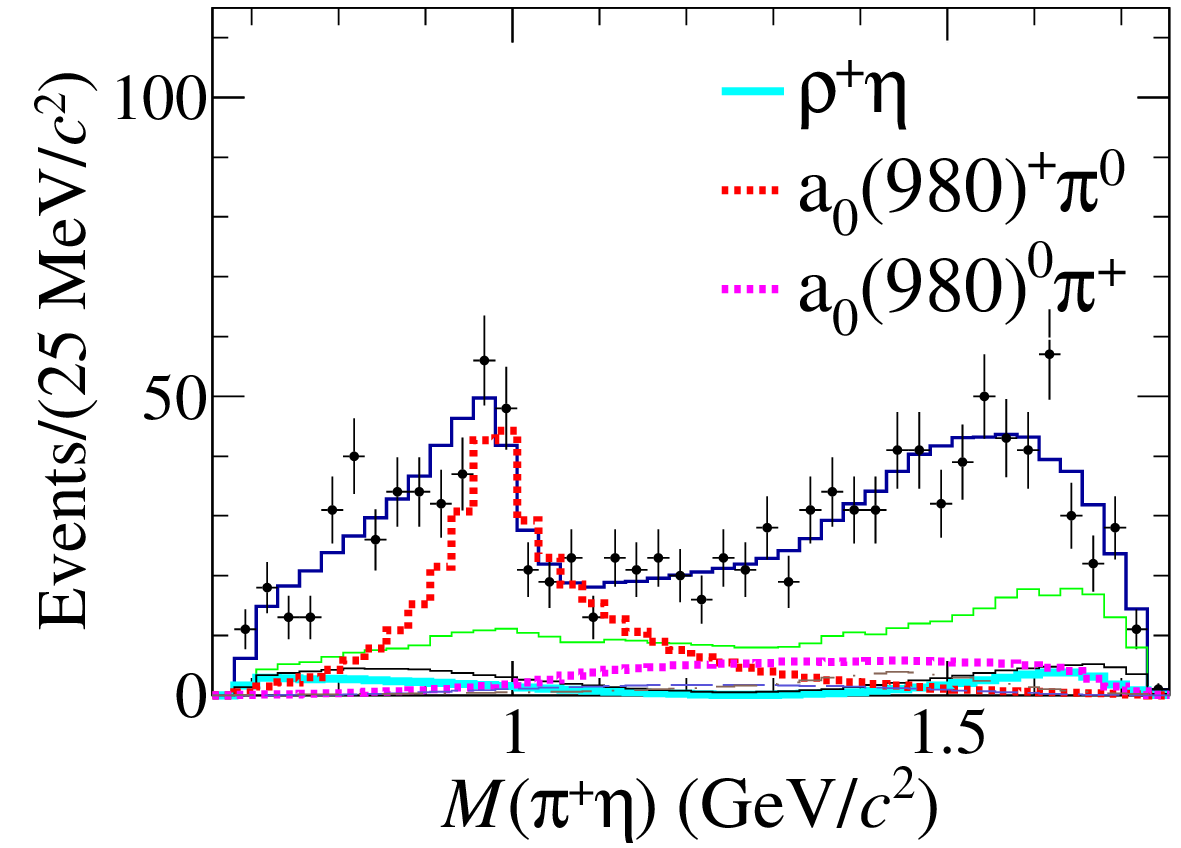}
\put(-85,65){(f)}
\end{minipage}
\begin{minipage}[b]{0.23\textwidth}
\epsfig{width=0.98\textwidth,clip=true,file=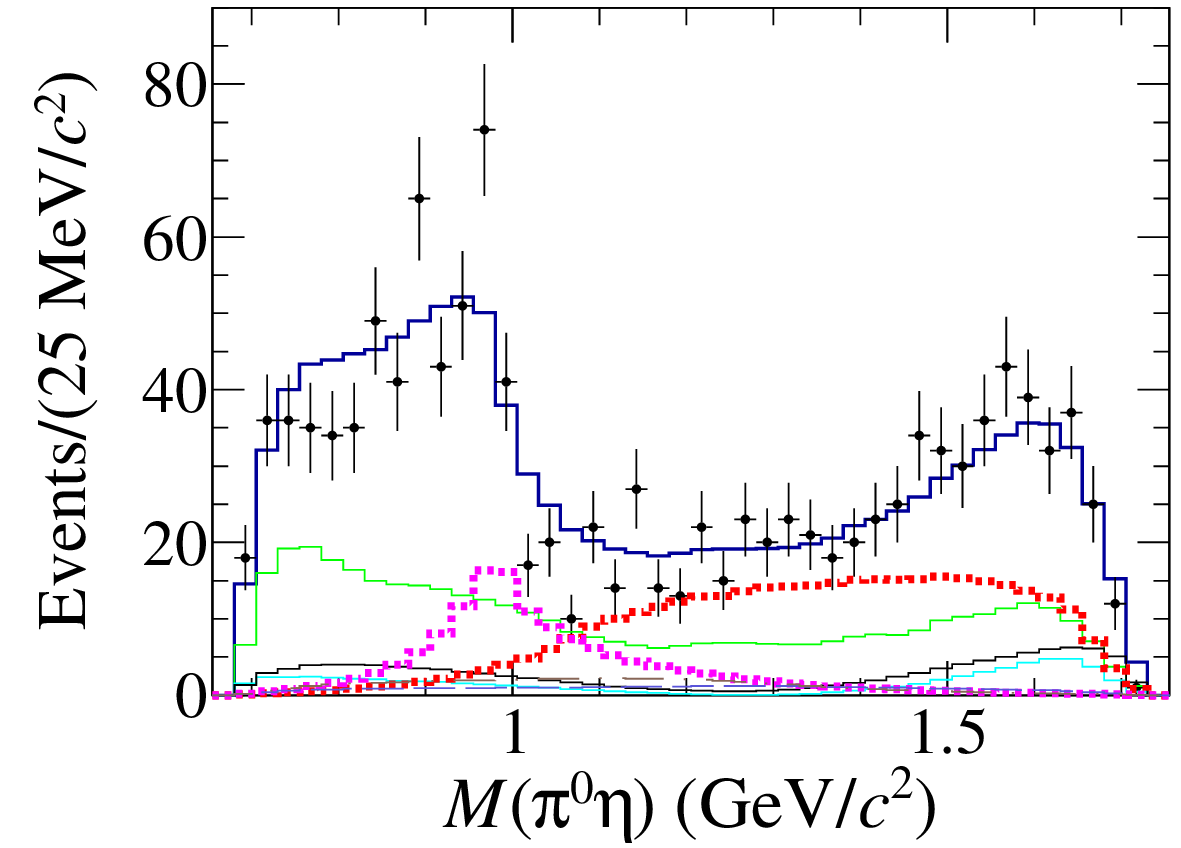}
\put(-85,65){(g)}
\end{minipage}
\begin{minipage}[b]{0.23\textwidth}
\epsfig{width=0.98\textwidth,clip=true,file=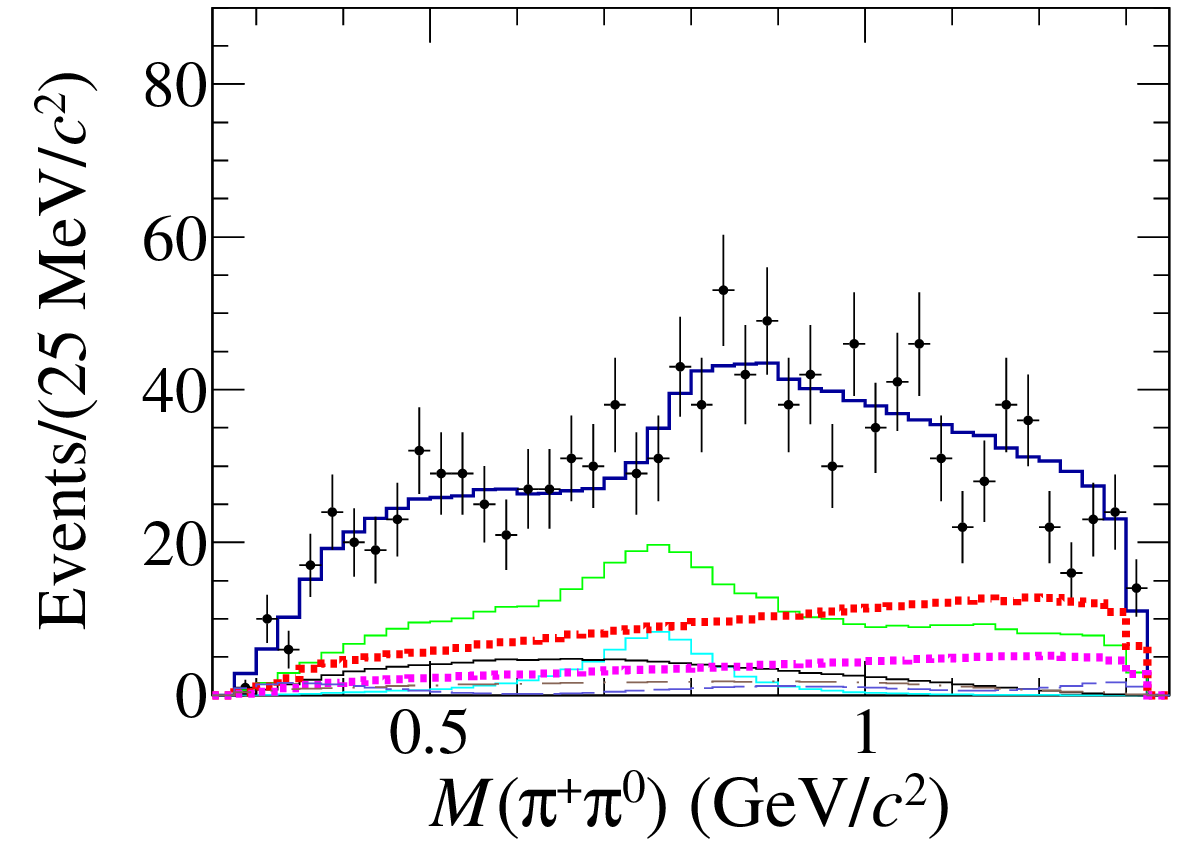}
\put(-85,65){(h)}
\end{minipage}
\caption{The Dalitz plot (a), the projections on $M(\pi^{+}\eta)$ (b),  $M(\pi^{-}\eta)$ (c),  $M(\pi^{+}\pi^{-})$  (d) the for $D^{0} \to \pi^{+}\pi^{-}\eta$ channel and
 the Dalitz plot (e), the projections on  $M(\pi^{+}\eta)$ (f),  $M(\pi^{0}\eta)$(g), $M(\pi^{+}\pi^{0})$ (h) for the $D^{+} \to \pi^{+}\pi^{0}\eta$ channel.
In the projections, the dots with error bars are data, the blue lines are the fit curves and the green histograms are the backgrounds; the cyan solid, pink dashed and red dashed lines are the contributions from the intermediate states  $\rho^{0(+)}$, $a_{0}(980)^{-(0)}$ and $a_{0}(980)^{+}$, respectively.}
\label{fig:projection}
\end{center}
\end{figure}

In the data projections,  a significant $\rho^{0}$ peak appears for the $D^{0}$ channel, 
while for the $D^{+}$ channel there is no evident $\rho^{+}$ peak but a significant $a_{0}(980)^{+}$ peak is observed. 
Therefore, the decays $D^{0} \to \rho^{0} \eta$ and $D^{+} \to a_{0}(980)^{+} \pi^{0}$ are chosen as the reference amplitudes for the $D^{0}$ and $D^{+}$ channels, respectively.
All contributions with significance larger than $3\sigma$ are retained for further analysis. 
Here, the significance is calculated using the changes of $\ln L$ and the number of degrees of freedom (NDF) 
when the fit is performed with and without the corresponding amplitude included. Following this criterion, 
six intermediate states are retained in the fit model for both channels.
The decay amplitudes and the corresponding $\phi_{\alpha}$, $\mathrm{FF}_{\alpha}$, significance, BF, and $r_{+/-(0)}$ values are listed in Table~\ref{tab:res}.
\begin{table*}[htbp]
\centering
\caption{The phases, FFs, statistical significances and BFs for various amplitudes.
The first and second uncertainties are statistical and systematic, respectively. 
The intermediate states are reconstructed in the decays $\rho \to \pi \pi$, $a_{0} \to \pi \eta$ and $a_{2} \to \pi \eta$.
}
\begin{threeparttable}
\begin{tabular}{l|cccc} \hline
Amplitude                            &  Phase (in unit rad)                 & FF (\%)                  & Significance ($\sigma$)           & BF ($\times 10^{-3}$) \\ \hline
$D^{0} \to \rho^{0}\eta$             &  $0$ (fixed)             & $15.2\pm1.7\pm1.0$       & $>10$                   & $0.19\pm0.02\pm0.01$              \\
$D^{0} \to a_{0}(980)^{-}\pi^{+}$    & ~~\,$0.06\pm0.16\pm0.12$    &  ~\,$5.9\pm1.3\pm1.0$ & ~~~\,8.9                        & $0.07\pm0.02\pm0.01$              \\
$D^{0} \to a_{0}(980)^{+}\pi^{-}$    & $-1.06\pm0.12\pm0.10$    & $44.0\pm4.0\pm5.3$       & $>10$                      & $0.55\pm0.05\pm0.07$              \\
$D^{0} \to a_{2}(1320)^{+}\pi^{-}$   & $-1.16\pm0.25\pm0.23$    &  ~\,$2.1\pm0.9\pm0.8$    & ~~~\,4.5                        & $0.03\pm0.01\pm0.01$              \\
$D^{0} \to a_{2}(1700)^{+}\pi^{-}$   &  ~~\,$0.08\pm0.17\pm0.23$    &  ~\,$5.5\pm1.8\pm2.7$ & ~~~\,6.1                        & $0.07\pm0.02\pm0.03$              \\
$D^{0} \to (\pi^{+}\pi^{-})_{S-\mathrm{wave}} \eta$& $-0.92\pm0.29\pm0.14$ &  ~\,$3.9\pm1.8\pm2.1$ & ~~~\,5.3          & $0.05\pm0.02\pm0.03$              \\
$r_{+/-}$                            &                          &$7.5^{+2.5}_{-0.8}\pm1.7$& ~~~\,7.7\tnote{*}  & -   \\
\hline
$D^{+} \to \rho^{+}\eta$              &$-4.03\pm0.19\pm0.13$  & ~\,$ 9.3\pm3.0\pm2.1$      & ~~~\,6.0               & $0.20\pm0.07\pm0.05$\\
$D^{+} \to (\pi^{+}\pi^{0})_{V} \eta$ &$-0.64\pm0.22\pm0.19$  & $15.8\pm4.8\pm5.2$      & ~~~\,4.7               & $0.34\pm0.11\pm0.11$\\
$D^{+} \to a_{0}(980)^{+}\pi^{0}$     & ~~\,$0$ (fixed)           & $43.7\pm5.6\pm1.9$      & ~~~\,9.1               & $0.95\pm0.12\pm0.05$\\
$D^{+} \to a_{0}(980)^{0}\pi^{+}$     &$ ~~\,2.44\pm0.20\pm0.10$  & $17.0\pm4.4\pm1.7$      & ~~~\,7.9               & $0.37\pm0.10\pm0.04$\\
$D^{+} \to a_{2}(1700)^{+}\pi^{0}$    &$ ~~\,0.92\pm0.20\pm0.14$  & ~\,$ 4.2\pm2.1\pm0.7$      & ~~~\,3.6               & $0.09\pm0.05\pm0.02$\\
$D^{+} \to a_{0}(1450)^{+}\pi^{0}$    &$ ~~\,0.63\pm0.41\pm0.30$  & ~\,$ 7.0\pm2.8\pm0.7$      & ~~~\,4.7               & $0.15\pm0.06\pm0.02$\\ 
$r_{+/0}$                             &                       & ~\,$2.6\pm0.6\pm0.3$    & ~~~\,4.0\tnote{*}  & - \\
\hline
\end{tabular}
\label{tab:res}
\begin{tablenotes}
    \footnotesize
    \item[*] The significance is for the test hypothesis $r=1.0$.
\end{tablenotes}
\end{threeparttable}
\end{table*}
The Dalitz plots and the projections are shown in Fig.~\ref{fig:projection}.
The fit quality is determined by calculating the $\chi^{2}$ of the fit using an adaptive binning of the $M^{2}(\pi^{+}\eta)$ versus $M^{2}(\pi^{-}\eta)$ ($M^{2}(\pi^{0}\eta)$) Dalitz plot with each bin containing at least 10 events. The resulting $\chi^{2}/\mathrm{NDF}$ is $136.6/138$ ($131.6/99$) for the $D^0 (D^+)$ channel.

In Table~\ref{tab:res}, the second uncertainties are systematic and arise from the following sources:
(I) the coupling with the $\pi\eta^{\prime}$ channel in the $a_{0}(980)$ line shape~\cite{BESIII:2016tqo}; 
(II)  the $a_{0}(980)^{\pm}$ mass and coupling constants, changed within the uncertainties given by Ref.~\cite{BESIII:2016tqo}; 
(III) the masses and widths of the $\rho^{0}$, $a_{2}(1320)^{+}$ and $a_{2}(1700)^{+}$ states, changed within the uncertainties given by PDG~\cite{Workman:2022ynf}; 
(IV) the parameters in the K-matrix formalism, changed within the statistical uncertainties given in Ref.~\cite{LHCb:2015klp} (only for the $D^{0}$ channel); 
(V) the effective radii for the intermediate resonances and for the $D^{0(+)}$ state, estimated by varying the effective radii by $\pm 1~\mathrm{GeV}^{-1}$; 
(VI) the background level, estimated from the uncertainty of the signal ratio in data;
(VII) the background shape, estimated by checking the effect on the fit results with background shapes extracted from the different sideband regions; 
(VIII) the fitter performance, estimated by fitting three hundred signal MC samples with the same size as the data sample.
The fits show good agreement between the fitted and the input values for the parameters in the amplitude model.
These systematic uncertainties are estimated separately by taking the difference between the values of $\phi_{\alpha}$, $\mathrm{FF}_{\alpha}$, $r_{+/-}$ and $r_{+/0}$ obtained by the alternative and the baseline fits. 
Since varying the propagators causes different normalization factors, only the effect on the $\mathrm{FF}_{\alpha}$ is considered~\cite{BESIII:2019jjr} from source (I) for the corresponding amplitudes, 
which is also similar for source (IV).
The total systematic uncertainties are obtained by adding each term in quadrature.

\begin{figure}[htbp]
\begin{center}
\begin{minipage}[b]{0.23\textwidth}
\epsfig{width=0.98\textwidth,clip=true,file=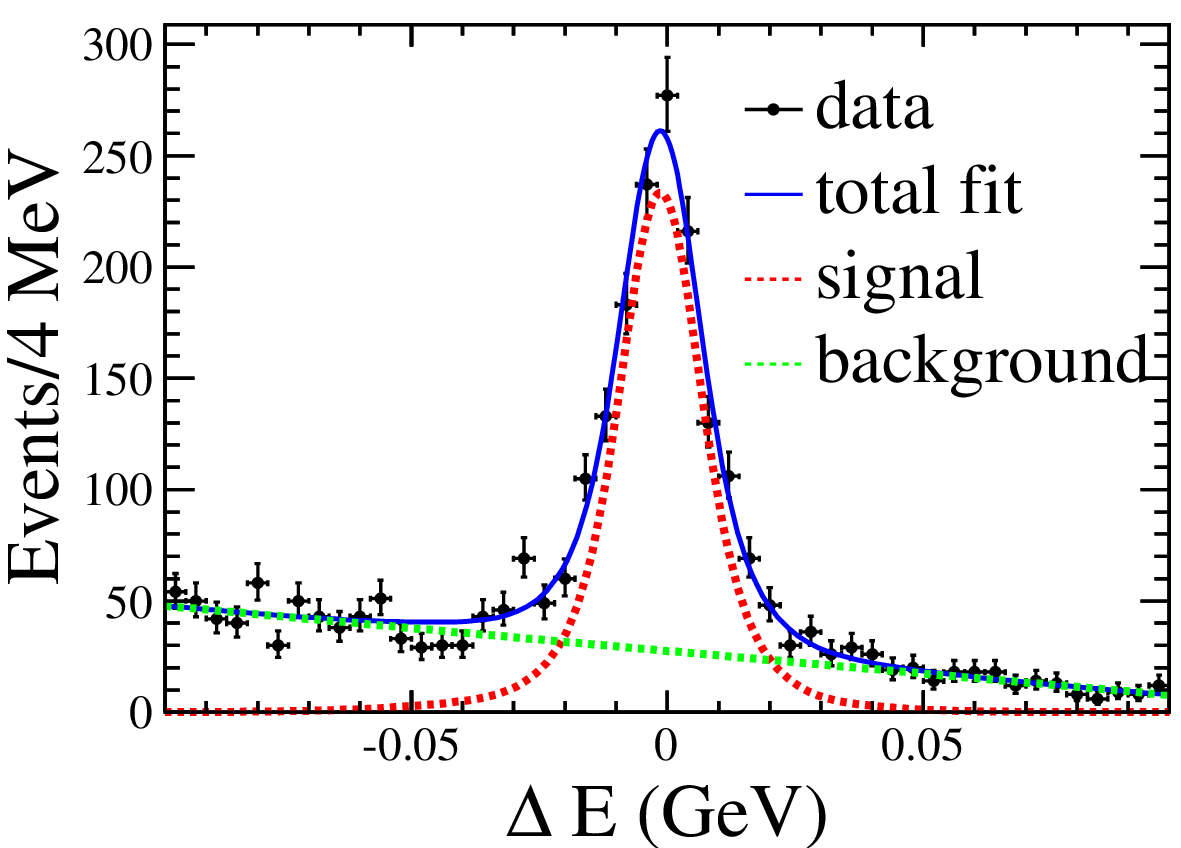}
\put(-80,55){(a)}
\end{minipage}
\begin{minipage}[b]{0.23\textwidth}
\epsfig{width=0.98\textwidth,clip=true,file=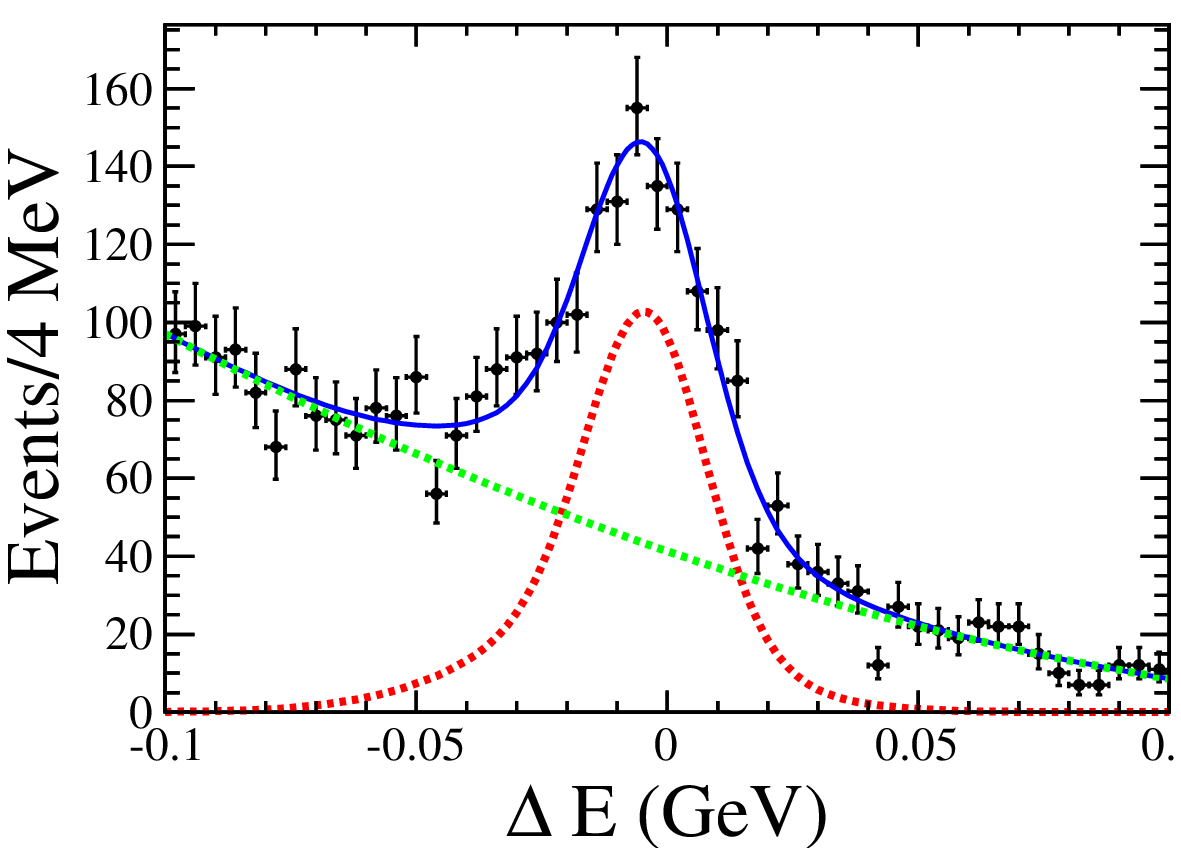}
\put(-80,55){(b)}
\end{minipage}
\caption{Fits to the $\Delta E$ distributions for (a) $D^{0} \to \pi^{+}\pi^{-}\eta$ and (b) $D^{+} \to \pi^{+}\pi^{0}\eta$. The dots with error bars are data.
The blue solid, red dashed and green dashed lines are the total fit, the signal and the background contribution, respectively.}
\label{fig:DT}
\end{center}
\end{figure}

The BFs of each sub-process in Table~\ref{tab:res} are calculated with $\mathcal{B}_{\alpha} = \mathrm{FF}_{\alpha} \times \mathcal{B}(D^{0(+)}\to \pi^{+} \pi^{-(0)} \eta)$. 
In the BF measurements, tighter windows than those used for the MVA selections, $0.505<M(\gamma\gamma)_{\eta}<0.570$~GeV$/c^{2}$ and $\chi^{2}(\eta)<50$, are used. 
The total decay BFs are measured with the DT method as $\mathcal{B} = Y_{\mathrm{DT}}/Y_{\mathrm{ST}}\epsilon_{\mathrm{sig}}\mathcal{B}_{\mathrm{sub}}$;
here, $Y_{\mathrm{ST}}$ is the total single tag (ST) yield, which is $(6897.1\pm8.2)\times10^{3}$ [$(4176.9\pm2.8)\times10^{3}$] for the $D^{0}$ ($D^{+}$) channel; $\mathcal{B}_{\mathrm{sub}}$ is the BF of $\pi^{0}(\eta) \to \gamma\gamma$;
$\epsilon_{\mathrm{sig}} $ is the weighted signal efficiency $\epsilon_{\mathrm{sig}} = \sum_{i}{\frac{Y^{(i)}_{\mathrm{ST}}}{\epsilon^{(i)}_{\mathrm{ST}}}\epsilon^{(i)}_{\mathrm{DT}}}/Y_{\mathrm{ST}}$, where 
the $Y^{(i)}_{\mathrm{ST}}$ and $\epsilon^{(i)}_{\mathrm{ST~(DT)}}$ are the ST yield and ST (DT) efficiencies for the $i^{\mathrm{th}}$ tag channels, respectively.
The DT yields $Y_{\mathrm{DT}}$ are extracted using a fit to the $\Delta E$ distributions without applying the signal window, as shown in Fig.~\ref{fig:DT}. In the fit, the signal PDF is parameterized as the sum 
of a bifurcated Gaussian~\cite{BifurcatedGauss} and a double Gaussian function, where the two functions have the same mean value. 
All the parameters except for the mean value are determined by the fit to the signal MC sample. 
The background function is described by a second-order Chebychev polynomial, validated by using the inclusive MC sample.
From the fits, we obtain $Y_{\mathrm{DT}}(D^{0}) = 1369\pm48$ and $Y_{\mathrm{DT}}(D^{+}) = 949\pm54$.

The total BFs of the $D^{0} \to \pi^{+} \pi^{-} \eta$ and $D^{+} \to \pi^{+} \pi^{0} \eta$ channels are measured to be
$(1.24\pm0.04_{\mathrm{stat.}}\pm0.03_{\mathrm{syst.}}) \times 10^{-3}$ and
$(2.18\pm0.12_{\mathrm{stat.}}\pm0.05_{\mathrm{syst.}}) \times 10^{-3}$, respectively. 
Here, the systematic uncertainties for the $D^{0}$ and $D^{+}$ channels include:
PID (1.0\% and 0.5\%); tracking efficiency (1.0\% and 0.5\%); 
$\eta/\pi^{0}$ reconstruction (0.8\% and 1.6\%), determined from hadronic DT $D\bar{D}$ events; 
signal shape (0.3\% and 0.1\%), estimated by the change of $Y_{\mathrm{\mathrm{DT}}}$ by altering the parameters in the signal shape within the uncertainties; 
background shape (0.6\% and 1.4\%), estimated by using a third-order Chebychev polynomial instead of the second-order one;    
$M_{\mathrm{BC}}$ window (0.3\% and 0.0\%), determined 
with the $D^{0} \to K^{-}\pi^{+}\eta$ control sample for the $D^{0}$ channel and negligible for the $D^{+}$ channel due to the loose window requirement;  
MC sample size (0.1\% and 0.1\%); 
quantum correlations (0.9\%, only for the $D^{0}$ channel), quoted from Ref.~\cite{BESIII:2019xhl}; 
fitter performance (0.3\% and 0.6\%), estimated from the inclusive MC sample; 
MC generator (0.2\% and 0.4\%), estimated by varying the input parameters in the generator according to the error matrix obtained from the fit to data; 
uncertainties from $\mathcal{B}(\eta/\pi^{0} \to \gamma \gamma)$ (0.5\% and 0.5\%), quoted from the PDG~\cite{Workman:2022ynf}.

From the amplitude model we calculate the ratios $r_{+/-} = 7.5^{+2.5}_{-0.8\,{\rm stat.}} \pm 1.7_{\rm syst.}$ and $r_{+/0} = 2.6 \pm 0.6_{\rm stat.}  \pm 0.3_{\rm syst.}$, 
which are both significantly higher than unity. Especially for the $D^{0}$ channel, the $r_{+/-}$ higher than the results of naive calculations that do not allow for 
enhancements to the WE contributions by two orders of magnitude~\cite{Cheng:2022vbw}.

In summary, we present the first amplitude analysis of the $D^{0(+)} \to \pi^{+} \pi^{-(0)} \eta$ channel.
The decays $D^{0(+)} \to a_{0}(980)^{+} \pi^{-(0)}$ and $D^{0(+)} \to a_{0}(980)^{-(0)} \pi^{+}$ are observed for the first time 
with statistical significances $>10\sigma$ ($9.1\sigma$) and $8.9\sigma$ ($7.9\sigma$), respectively.
The $a_{0}(980)^{+}$ is identified as the dominant intermediate resonance in both channels, and its contribution is found to be significantly larger than that of the 
$a_{0}(980)^{-(0)}$ state. 

For the $D^{+}$ channel, the measured value of $r_{+/0}$ indicates that the symmetry observed in $D_{s}^{+}$ decays is here violated.
Furthermore, in contrast with the large BF of $D_{s}^{+} \to \rho^{+}\eta$, the low BF of $D^{+} \to \rho^{+}\eta$ shows the importance of the $K^{*}K \to a_{0}(980)\pi$ re-scattering process. 
For the $D^{0}$ channel, the measured value of $r_{+/-}$ disagrees with theoretical predictions that ignore non-perturbative effects by two orders of magnitude.
Estimations suggest that the size for the amplitude of WE process is even larger than that of the T diagram~\cite{Cheng:2022vbw,npcontribution}.  
This dominance of the WE process in $D \to SP$ decays is to be contrasted with the situation in $D \to SP$, $D \to PP$ and $D \to VP$ ($V$ denotes vector particle) decays~\cite{Cheng:2022vbw,Cheng:2024hdo}, indicating the very important role that FSI plays in the $D \to SP$ sector.
In analogy to theoretical interpretations for WA process, the re-scattering contributions also suggest that we would expect $r_{+/-}<1$~\cite{c2is0P3}, in contradiction to our measurement. 
In addition, the resonance $a_{0}(1817)$ is not observed in both channels.
The measured BFs and ratios are highly valuable for improving the understanding of the role that the $a_{0}(980)$ plays in charm-meson decays and the theoretical interpretations of the nature of the  $a_{0}(980)$ state.

The authors greatly thank Professor H.Y. Cheng from Institute of Physics, Academia Sinica for the useful discussions.
The BESIII Collaboration thanks the staff of BEPCII and the IHEP computing center for their strong support. This work is supported in part by National Key R\&D Program of China under Contracts Nos. 2023YFA1606000, 2020YFA0406300, 2020YFA0406400; National Natural Science Foundation of China (NSFC) under Contracts Nos. 11635010, 11735014, 11935015, 11935016, 11935018, 11961141012, 12025502, 12035009, 12035013, 12061131003, 12192260, 12192261, 12192262, 12192263, 12192264, 12192265, 12221005, 12225509, 12235017, 12205384, 12361141819; the Chinese Academy of Sciences (CAS) Large-Scale Scientific Facility Program; the CAS Center for Excellence in Particle Physics (CCEPP); Joint Large-Scale Scientific Facility Funds of the NSFC and CAS under Contract No. U1832207, U2032104; 100 Talents Program of CAS; The Institute of Nuclear and Particle Physics (INPAC) and Shanghai Key Laboratory for Particle Physics and Cosmology; German Research Foundation DFG under Contracts Nos. 455635585, FOR5327, GRK 2149; Istituto Nazionale di Fisica Nucleare, Italy; Ministry of Development of Turkey under Contract No. DPT2006K-120470; National Research Foundation of Korea under Contract No. NRF-2022R1A2C1092335; National Science and Technology fund of Mongolia; National Science Research and Innovation Fund (NSRF) via the Program Management Unit for Human Resources \& Institutional Development, Research and Innovation of Thailand under Contract No. B16F640076; Polish National Science Centre under Contract No. 2019/35/O/ST2/02907; The Swedish Research Council; U. S. Department of Energy under Contract No. DE-FG02-05ER41374


\begin{thebibliography}{99}
\bibitem{BifurcatedGauss} The bifurcated Gaussian function is the altered Gaussian function with allowing the left width and right width being different.
\bibitem{npcontribution} The diagram for $D^{0} \to a_{0}(980)^{\pm}\pi^{\mp}$ consists of a T diagram and a WE diagram. 
To estimate the relative size of the WE amplitude over the T amplitude, 
we set the amplitude of the T diagram in $D^{0} \to a_{0}(980)^{-} \pi^{+}$ to be $1.0e^{i0.0}$, and denote that for the T diagram in $D^{0} \to a_{0}(980)^{+} \pi^{-}$ to be $be^{i\phi_{b}}$, 
those amplitudes for the WE diagrams for $a_{0}(980)^{-}$ and $a_{0}(980)^{+}$ involved sub-modes to be $xe^{i\phi_{-}}$ and $cxe^{i\phi_{+}}$, respectively, and write $r_{+/-}$ as $r$ for brevity.  
Here, the parameters $b$, $c$ and $x$ are real numbers. 
According to how the relative BF is determined, we obtain $\frac{|cxe^{i\phi_{+}} + be^{\phi_b}|^{2}}{|xe^{i\phi_{-}} + 1|^{2}} = r$. Then $(r-c^{2})x^{2} + 2[r\cos(\phi_{-}-\phi_{b}) - cb\cos(\phi_{+}-\phi_{b})]x + (r - b^{2}) = 0$; therefore, the minimum of $|x|$ is $\sqrt{\frac{r-b^{2}}{r-c^{2}}}$. The parameter $b^2$ is expected to be less than 0.05~\cite{Cheng:2022vbw}. The parameter $c$, the relative magnitude of the amplitude for the WE diagram in $D^{0} \to a_{0}(980)^{+} \pi^{-}$ over that in $D^{0} \to a_{0}(980)^{-} \pi^{+}$, can be estimated from the largest re-scattering contribution, which is $\frac{\mathcal{B}(D^{0} \to K^{+}K^{*-})}{\mathcal{B}(D^{0} \to K^{-}K^{*+})}$~\cite{c2is0P3}, therefore $c^{2}\sim0.3$. It follows that we can expect that $b^{2}<c^{2}$, and so $|x|>1$.
\bibitem{c2is0P3} In analogy to theoretical interpretations for WA process, re-scattering through $\rho\eta \to a_{0}(980)\pi$
and $K^{*}K \to a_{0}(980)\pi$ triangle diagrams can also be invoked to explain the large FSI in the WE process. 
However, since the BFs of $D^{0} \to \rho^{0}\eta$, $D^{0} \to K^{*0}\bar{K}^{0}$ and $D^{0} \to \bar{K}^{*0}K^{0}$~\cite{Workman:2022ynf} are small, 
the large FSI contribution is then expected from the re-scattering of $K^{*\pm}K^{\mp} \to a_{0}(980)^{\mp}\pi^{\pm}$. 
While through the triangle diagram, only the re-scatterings of $K^- K^{*+} \to \pi^{+} a_{0}(980)^{-}$ and $K^+K^{*-} \to \pi^- a_{0}(980)^+$ are allowed.
Therefore if the WE contribution is dominant, 
$r_{+/-} \sim \frac{\mathcal{B}(D^{0} \to K^{+}K^{*-})}{\mathcal{B}(D^{0} \to K^{-}K^{*+})}$, which is only about 0.3.
\end{thebibliography}
\end{document}